# Multispectral Polarization-Insensitive Graphene/Silicon Guided Mode Resonance Active Metasurfaces

**Prateeksha Sharma[1,2], Dor Oz[2], Eleftheria Lampadariou[3], Spyros Doukas[3], Elefterios Lidorikis[3] and Ilya Goykhman[2*]**

[1]*Faculty of Electrical and Computer Engineering, Technion, Haifa 320000, Israel*
[2]*Institute of Applied Physics, The Faculty of Science and The Center for Nanoscience and Nanotechnology, The Hebrew University of Jerusalem, Jerusalem 91904, Israel.*
[3]*Department of Materials Science and Engineering, University of Ioannina, 45110, Ioannina, Greece*

**Abstract:** We investigate advanced CMOS-compatible Graphene/Silicon active metasurfaces based on guided-mode resonance filters. The simulated results show a high extinction ratio (>25 dB), narrow linewidth (~1.5 nm @1550 nm), quality factor of Q~1000, and polarization-insensitive operation. By taking advantage of graphene broadband absorption, we present a multispectral operation in the MIR using simple geometrical scaling rules. We further showcase that the same device architecture can be employed for thermo-optic tuning using graphene as an integrated microheater. Our work contributes to the development of advanced broadband silicon-based active metasurfaces for tunable spectral filters and laser mirrors, optical switches, modulators, and sensors.

## 1. Introduction

Optical metasurfaces are a class of artificially engineered materials that offer unique ways of manipulating the electromagnetic wave characteristics at the nanoscale [1]. These two-dimensional structures commonly comprise arrays of subwavelength antennas, resonators, or gratings designed to dynamically control the phase, amplitude, and polarization of transmitted, reflected, and diffracted light [1-3]. Introducing active functionalities to optical metasurfaces opens new avenues in optics, telecommunication, imaging, and sensing, [3-5] enabling next generation flat optical devices, such as tunable lenses and mirrors [6], reconfigurable antennas [7], holographic displays [8], beam steering, beamforming and frequency tuning [9], bio-imaging [10] and sensing for environmental monitoring and security [11]. Thus, the efficient integration of novel optical materials, whose optical properties can modulate in response to external stimuli such as electric field [12], heating [13], or optical pumping [14], is of great interest for the development of advanced active metasurfaces.

Over the years, different material systems, including phase change materials [15], transparent conductive oxides [16], liquid crystals [17], III-V semiconductors [18], and recently two-dimensional (2D) materials [19], have been integrated with metasurfaces to provide active functionalities. Specifically, liquid crystals [20] and phase change materials [21] have been widely used to modulate the phase of the scattered light, but their limited operation speed (< 1 kHz) sets a technical bottleneck for the widespread use of these devices [20, 21]. III-V semiconductor quantum-wells based metasurfaces have demonstrated faster response (~100 MHz) for phase modulation [22, 23] but showed only a limited modulation depth ($\Delta\varphi < \pi/2$), which is less suitable for practical applications [22, 23]. Transparent conducting oxides like indium tin oxide (ITO) exhibit both optical transparency and electrical conductivity, with tunable dielectric permittivity and enhanced light-matter interaction at the epsilon-near-zero (ENZ) regime [24]. The latter is particularly interesting since it provides higher modulation efficiency ($\Delta\varphi > \pi$) and operation speed (>10 MHz) [25]. However, the ITO fragility, leakage current limitations, scarcity, and cost, restrain its potential for extensive long-term employment. On the other hand, single-layer graphene (SLG), a 2D semimetal, offers a unique combination of ultra-broadband spectral response [26], electrical tunability of complex optical refractive index from visible-to-terahertz (VIS-THz) wavelengths [27], high carrier mobility and operation speed [28], and importantly compatibility with mature silicon technology and back-end-of-line (BEOL) complementary metal-oxide-semiconductor (CMOS) fabrication process [29]. Due to its 2D nature, external voltage bias can efficiently modulate graphene's free carriers' concentration and provide a sufficient Fermi-level shift to change the optical loss (Pauli blocking) and the SLG refractive index [28]. Graphene-integrated active metasurfaces based on free carriers' effect have been demonstrated with graphene micro ribbons [30], split-ring resonators [31], antennas [32], and plasmonic metamaterials [33] operating in infrared (IR), and terahertz (THz) spectra. The broadband spectral response, efficient optical modulation across multiple spectral bands, and CMOS compatibility make graphene an attractive candidate for implementation in silicon-based, multispectral active metasurfaces.

In this work, we propose and investigate advanced designs of CMOS-compatible graphene-silicon guided-mode resonance (GMR) based active metasurfaces, which exhibit electrically controlled and switchable reflection, high extinction ratio (~28 dB), and spectral tunability. Our designs can be tailored to polarization-insensitive operation using 2D grating architectures. We exploit the broadband capabilities of electrically tunable optical absorption in graphene and demonstrate a multispectral device operation from short-wave to mid-wave infrared wavelengths through geometric scaling while maintaining performance metrics, such as high extinction ratio (>20dB), quality factor, and narrow resonance linewidth. Beyond the electro-optic tuning based on graphene complex refractive index modulation, we showcase that the same device architecture can be employed for the thermo-optic tuning using graphene as an integrated microheater. This dual graphene functionality can provide an additional degree of freedom to control the GMR filter spectral response and compensate for fabrication tolerances when the device is integrated into more complex photonic systems and functionalities.

## 2. Tuning by Electro-Optic Effect

### 2.1 Design Considerations

Figure 1a shows the materials stack-up of CMOS-compatible electrically tunable graphene/Si metasurfaces implemented by GMR spectral filters. GMR devices have emerged as promising candidates for the realization of the metasurfaces and flat optical components for a variety of optoelectronic applications, including narrow-band filters [34], absorbers [35], optical switches [36], polarizers [37], modulators [36] and laser mirrors [38]. The basic structure of the GMR filter consists of a subwavelength grating integrated over a high-index waveguide medium [39, 40]. In this configuration, incident light couples to the leaky waveguide mode when the phase-matching conditions are satisfied at resonance. Spectral filtering is achieved by the interference between the incident wave and the out-diffracted leaky waveguide mode [39, 40]. Modern GMR filters [40] offer a high extinction ratio (ER) with almost 100% out-of-band rejection efficiency [40], where the resonance wavelength can be controlled by device geometry (grating period, filling factor, waveguide thickness), material properties (complex refractive index), incident illumination angel and light polarization [40].

In our study, we investigated two designs, namely polarization-sensitive and polarization-insensitive configurations (Figs. 1b,c), in which SLG is integrated onto a technologically relevant silicon-on-insulator (SOI) substrate with a buried oxide (BOX) of $t_{BOX} = 3$ μm and silicon device layer of $t_{Si} = 220$ nm. In both designs, a low refractive index dielectric grating (e.g., oxide blocks) with a period of $L = 800$ nm is employed to optimize the resonance condition at the telecom wavelengths $1.5 - 1.6$ μm and maximize the modulation extinction ratio. For polarization-sensitive configuration (Fig. 1b), a 1D grating with a width $W = 400$ nm, height $h = 90$ nm, and a 50% coverage $f = W/L = 0.5$ is used. Conversely, for the polarization-insensitive filter, a square lattice with $f = 0.6$ coverage, grating dimensions of $W = W_x = W_y = 480$ nm and $h = 190$ nm is implemented (Fig. 1c). A high-k hafnium oxide gate-dielectric layer (hafnia HfO$_2$, $t_{HfO_2} = 30$ nm) with a dielectric permittivity of $\varepsilon_{HfO_2} \sim 22$ [41] is considered for both configurations to allow substantial doping of the SLG (i.e. Fermi level $E_F \sim 0.6$ eV) using gate voltages $V_G < 6$ V.

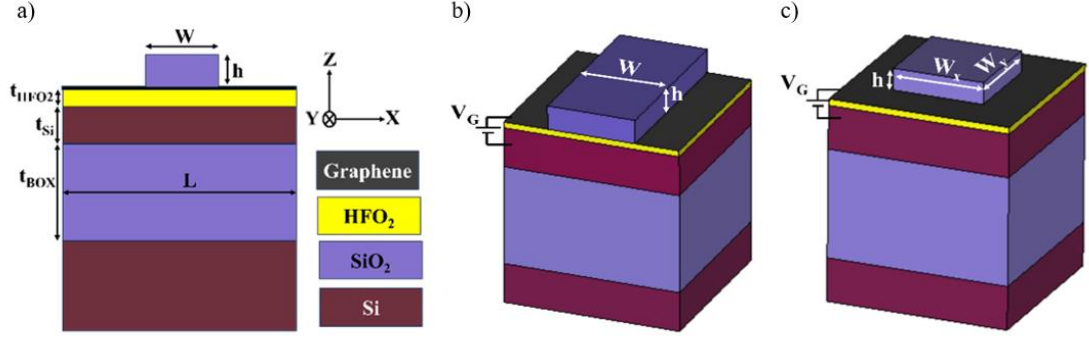

Fig. 1. (a) Cross-section and material stack-up view of 1D and 2D gratings. (b) Perspective view of polarization-sensitive configuration with 1D grating having $W = 400\ nm$ and $h = 90$ nm (c) Perspective view of polarization-insensitive configuration with 2D grating having $W = W_x = W_x = 480$ nm and $h = 190$ nm. In both configurations, the gratings period is $L = 800$ nm and hafnia thickness is $t_{\text{HfO}_2} = 30$ nm. The SLG graphene scattering time is $\tau = 0.2$ ps. The buried oxide thickness is $t_{\text{BOX}} = 3$ μm and the silicon device layer is $t_{\text{Si}} = 220$ nm.

We used the refractive index of silicon $n_{\text{Si}} \sim 3.48$ [42] and of silicon dioxide $n_{\text{SiO}_2} \sim 1.44$ [43] at the telecom wavelengths. The optical properties of SLG were calculated from its optical conductivity $\sigma_G$ which accounts for both the inter-band and intra-band transitions [44, 45]:

$$\sigma_G = \sigma_0 \left\{ \frac{1}{2}\left[\tanh\left(\frac{\hbar\omega+2\mu_c}{4k_bT}\right) + \tanh\left(\frac{\hbar\omega-2\mu_c}{4k_bT}\right)\right] - \frac{i}{2\pi}\log\left[\frac{(\hbar\omega+2\mu_c)^2}{(\hbar\omega-2\mu_c)^2+4(k_bT)^2}\right] + \frac{4i}{\pi}\frac{\mu_c}{\hbar\omega+i\hbar\gamma} \right\} \quad (1)$$

where $T$ is the temperature, $\mu_c$ is the chemical potential, $\omega$ is the angular frequency, $k_B$ is the Boltzmann constant, $\hbar$ is the reduced Planck constant, $\sigma_0 = q^2/(4\hbar)$ is the universal conductivity of undoped graphene [44, 45], $\gamma$ is the scattering rate, $\hbar\gamma = 1.65$ meV is the electron relaxation energy which corresponds to a relaxation time of ~0.2 ps of polycrystalline graphene with mobility of $\mu \sim 2000\ \text{cm}^2/\text{V}\cdot\text{s}$, and $q$ is the elementary electron charge [44, 45]. The in-plane dielectric permittivity of SLG is related to its surface conductivity and is given by [46]:

$$\varepsilon_G = 1 + \frac{i\sigma_G}{\omega\varepsilon_0 t_G} \quad (2)$$

where $\varepsilon_0$ is the dielectric permittivity of the vacuum, and $t_G = 0.34$ nm is the thickness of the SLG [47, 48]. Further, the refractive index can be calculated as:

$$n = \sqrt{\varepsilon_G} \quad (3)$$

From equations (1)-(3), the optical properties of SLG strongly depend on its chemical potential (i.e., Fermi level), allowing the realization of active voltage-tuned optical metasurfaces by integrating graphene in GMR filter configuration. Specifically, in our studies, SLG is separated from Si by a thin ($t_{\text{HfO}_2} = 30$ nm) high-k dielectric layer, forming a SLG/HfO$_2$/Si capacitor, to enable electrostatic doping of the SLG by applying a gate voltage $V_G$. The latter modulates the free carriers' density $n_G$ in graphene, shifting the Fermi level according to $E_F = \hbar v_F \sqrt{\pi n_G}$ [49], where $qn_G = V_G\, \varepsilon_0 \varepsilon_{\text{HfO}_2}/t_{\text{HfO}_2}$ is the surface charge per unit area and $v_F$ is the Fermi velocity of charge carriers in graphene. It should be noted in this context that the applied $V_G$ should operate the device below the breakdown field $E_{BD} = V_G/t_{\text{HfO}_2}$ of the gate dielectric. Given $E_{BD} \sim 20$ MV/cm [41] for hafnia, we considered for practical reasons the maximal induced SLG doping of $E_F \sim 0.6$ eV, well below the $E_{BD}$.

## 2.2 Results and Discussion

The device dimensions and material properties were optimized to obtain sharp resonances in the infrared spectra, having high reflectance (>90 %), narrow linewidth (~ 1 nm), and optimal extinction ratio upon electro-optic modulation induced by $V_G$. The finite-element optical simulations were conducted in a commercial software package CST Microwave Studio [50], where the graphene-integrated GMR unit cell (Fig. 1) was simulated with periodic boundary conditions in the lateral (i.e. $x - y$) directions and perfect-matched layer (PML) boundary conditions in vertical (i.e., $z$) direction. Figure 2 shows the simulated reflection spectra for 1D and 2D GMR-based metasurfaces with realistic ($E_F < 0.6$ eV) SLG doping induced by the SLG/HfO$_2$/Si capacitor, safely operated below the breakdown. The 1D-grating device (Fig. 1b) is polarization-sensitive; hence the reflection spectra are plotted for the transverse-magnetic (TM) mode only, as shown in Fig. 2a. The 2D square grating device (Fig. 1c) is polarization-insensitive, supporting both transverse-electric (TE) and TM modes, showing the same spectral response for both polarizations (Fig. 2c).

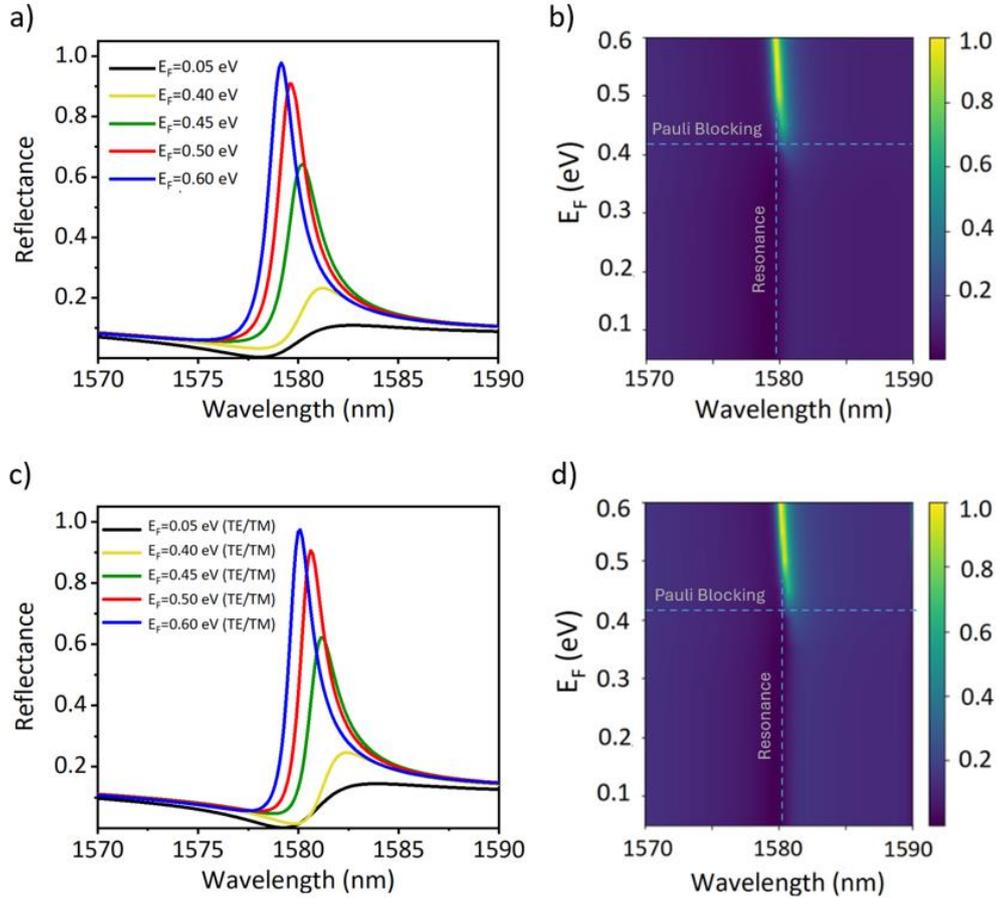

Fig. 2. Reflectance spectra and reflectance map at different Fermi levels for (a-b) TM polarized 1D grating with $W = 400$ nm, $h = 90$ nm, $f = 0.5$ and (c-d) polarization-insensitive 2D grating with $W_x = W_y = 480$ nm, $h = 190$ nm, $f = 0.6$. Other parameters for both gratings are set as $L = 0.8$ μm, $t_{HfO_2} = 30$ nm, $t_{Si} = 220$ nm, $t_{BOX} = 3$ μm and $\tau = 0.2$ ps.

The simulation results highlight that changing SLG doping by electrostatic gating can dynamically tune the reflectance and resonance wavelength. Specifically, for high doping levels, $E_F = 0.6$ eV, when graphene is in the Pauli blocking (transparency) regime, the reflectance values at resonance reach $R_{ON} > 95\%$. While for lower doping $E_F < 0.4$ eV, when SLG absorption is governed primarily by the interband transitions (Eq. 1), the reflectance drops below $R_{OFF} < 2\%$, showing the extinction ratio of ER = $10\log(R_{ON}/R_{OFF})$ >18 dB. The spectral response of 1D and 2D grating devices can be further adjusted by varying geometric parameters such as grating heights $h$ and coverage $f$, as presented in Fig. 3. In both configurations, the resonance wavelength is red-shifted when $h$ and $f$ increase. Thicker $h$ leads to higher reflectance but at the expense of broader resonance linewidth (Fig. 3a, c). The ER can be further optimized by using different coverage areas (Fig. 3b, d), leading to the optimal device parameters that were employed in our simulations, i.e. $h = 90$ nm, $f = 0.5$ and $h = 190$ nm, $f = 0.6$ for 1D and 2D gratings respectively.

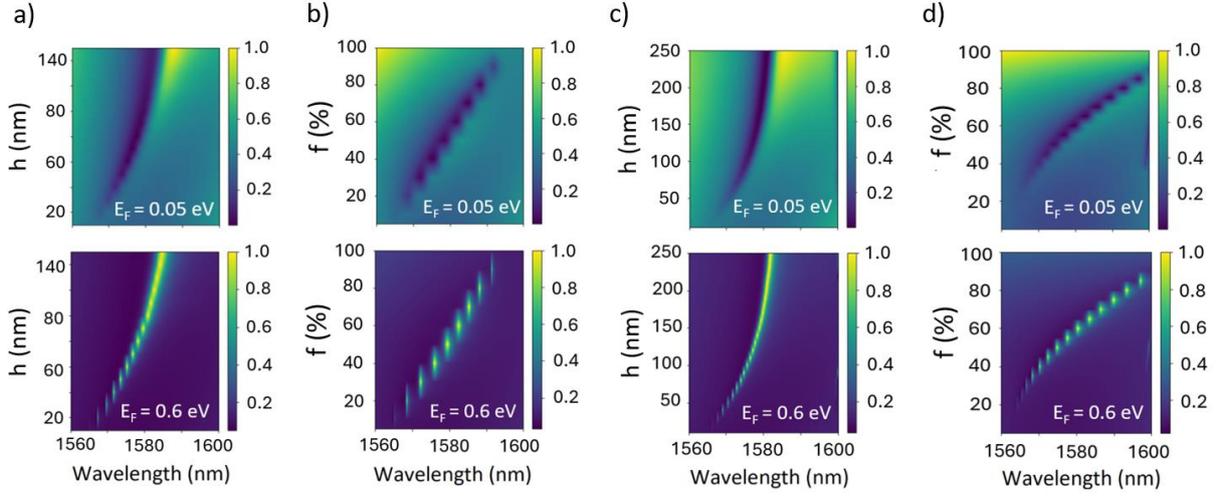

**Fig. 3**. Parametric studies of reflectance spectra of 1D and 2D devices at different doping levels, $R_{ON}$ ($E_F = 0.6$ eV) and $R_{OFF}$ ($E_F = 0.05$ eV). The reflectance map as a function of grating height (a) and (c), and as a function of surface coverage(c) and (d) for different $E_F$. The optimized parameters to maximize the extinction ratio are: $h = 90$ nm, $f = 0.5$ for 1D and $h = 190$ nm, $f = 0.6$ for 2D grating. Other parameters for both devices are set as $L = 0.8$ μm, $t_{HfO_2} = 30$ nm, $t_{Si} = 220$ nm, $t_{BOX} = 3$ μm and $\tau = 0.2$ ps.

The electric field (E-field) plots provide more evidence of the polarization-sensitive (insensitive) behavior, as shown in Fig. 4. The 1D grating configuration does not support TE-polarization, and the optical mode leaks into the substrate (Fig. 4a). On the contrary, the TM mode is coupled and guided in the Si/HfO$_2$/SLG heterostructure, enhancing light interaction with graphene (Fig. 4b). Similarly, the guided mode characteristics are observed for TE and TM modes in the polarization-insensitive 2D grating configuration, as shown in Figs. 4c,d.

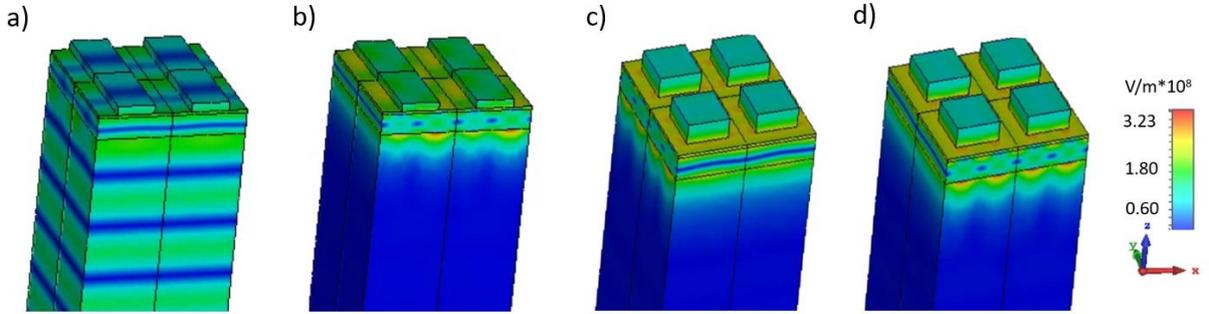

**Fig. 4**. E-field distribution at resonance ($E_F = 0.6$ eV) for 1D and 2D grating-based GMR metasurfaces. (a) TE-polarized 1D configuration. The optical mode leaks into the substrate. (b) TM-polarized 1D configuration. The optical mode is coupled to the hybrid Si/HfO$_2$/SLG waveguide heterostructure. (c) and (d) show the 2D configuration for TE- and TM-polarized illumination. The light is guided in Si/HfO$_2$/SLG heterostructure for TE- and TM-polarizations. The geometrical parameters used in the simulations are: $h = 90$ nm, $f = 0.5$ for 1D and $h = 190$ nm, $f = 0.6$ for 2D grating and for both devices $L = 0.8$ μm, $t_{HfO_2} = 30$ nm, $t_{Si} = 220$ nm, $t_{BOX} = 3$ μm and $\tau = 0.2$ ps.

The device reflectance is controlled by changing the SLG doping (Fig. 5a), which affects both the real and imaginary parts ($\sigma_{Re}$ and $\sigma_{Im}$) of optical conductivity and modulate graphene complex refractive index. The resonance wavelength is red-shifted when the Fermi level (doping) increases, due to the larger $\sigma_{Im}$ and the real part of the effective refractive index (Fig. 5b). On the other hand, the device reflectance values depend on optical loss in the SLG, i.e. $\sigma_{Re}$ (Eq. 2), bringing graphene into transparency when $E_F$ reaches the Pauli blocking regime ($E_F > \frac{E_{ph}}{2}$, where $E_{ph}$ is the photon energy). For instance, a substantial drop from 95% to 10% in reflectance is achieved when graphene is switched from transparency ($E_F > 0.5$ eV @1550 nm), where the interband absorption is suppressed, to the lossy regime ($E_F < 0.4$ eV) below the Pauli blocking (Fig. 5a). The ER of modulation was calculated for different doping levels with respect to the resonance at $E_F = 0.6$ eV, i.e. ER = $10\log(R_{0.6eV}/R_{E_F})$, as shown in Fig. 5c. It follows the trend of the real part of graphene permittivity $\varepsilon_{real}$ having a resonance at the Pauli blocking condition $E_{ph} = 2E_F$. Specifically, the ER first increases with the Fermi level, up to Pauli blocking ($E_F = 0.4$ eV @ 1550 nm), and then decreases in the transparency region. The material resonance is reproduced in the optical domain by tuning the graphene-activated GMR filter into a critical coupling condition, where two Fano-coupled oscillators (Fig. 2) of the guided and scattered light are critically coupled.

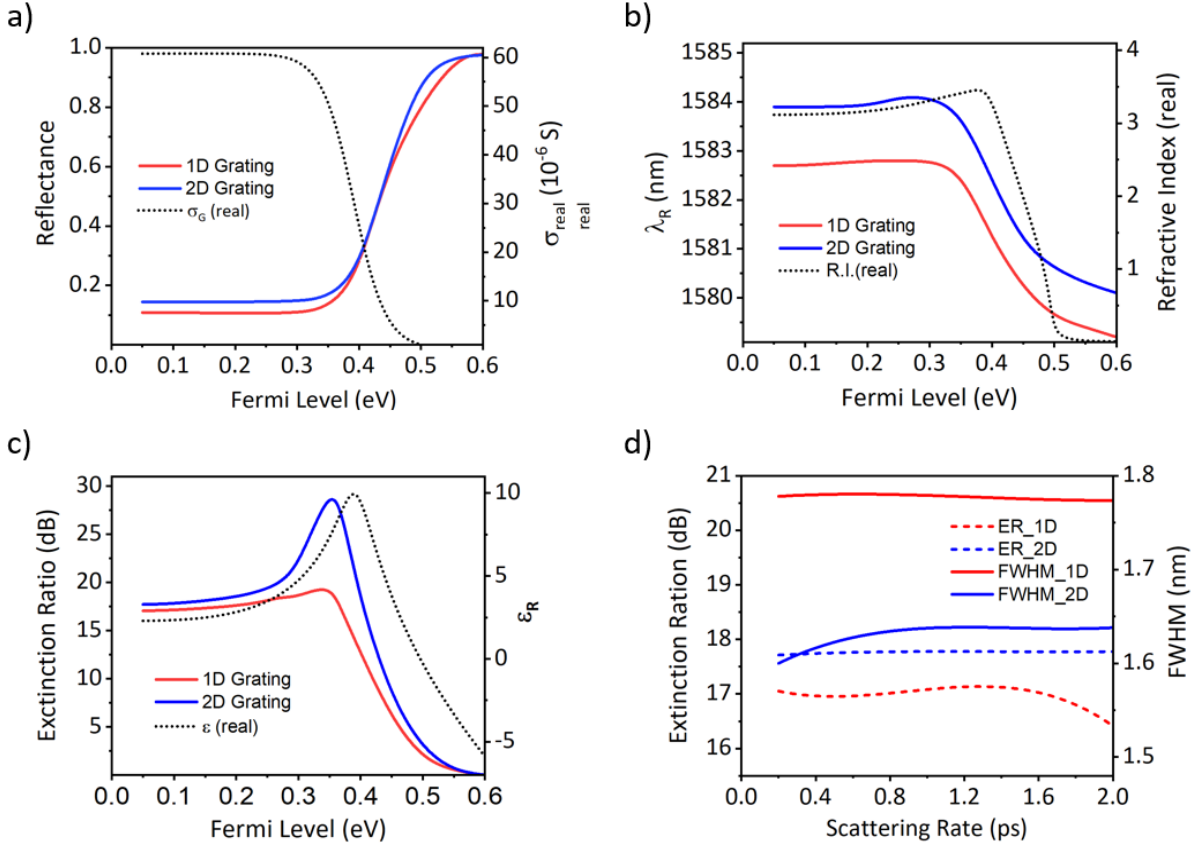

**Fig. 5.** (a) The device reflectance at resonance wavelength (red, blue lines) and a real part of graphene optical conductivity (dotted line) as a function of Fermi level. (b) Modulation of the resonance wavelength (red, blue lines) and graphene real part of refractive index (dashed line) as a function of Fermi level. (c) Extinction ratio ER = $10\log(R_{0.6eV}/R_{E_F})$ of modulation with respect to the resonance at $E_F = 0.6$ eV (red, blue lines) and graphene dielectric permittivity (dotted line) as a function of Fermi level. (d) Extinction ratio ER = $10\log(R_{0.6eV}/R_{0.05eV})$ at the reference wavelength of $E_F = 0.6$ eV (solid lines) and the resonance linewidth of full width at half maximum (FWHM, dashed line) as a function of graphene quality (i.e. scattering rate).

The ER of ~18 dB and ~28 dB for 1D and 2D gratings respectively were calculated for Fermi level modulation between $E_F = 0.6$ eV and $E_F = 0.35$ eV (Fig. 5c). The obtained ER and the resonance linewidth show insignificant variation for different scattering time $\tau$, as shown in Fig. 5d. Here, the ER was calculated between $E_F = 0.6$ eV in transparency and $E_F = 0.05$ eV in lossy regimes. As evident, the proposed devices offer high ER, sharp resonances with full-width at half-maximum (FWHM) of ~1.7 nm, and quality factor of $Q$~930 at $E_F = 0.6$ eV - almost independent of graphene scattering rate, which allows the device realization with graphene of different quality.

## 3. Scaling and Analysis at Mid-IR

SLG shows ultra-broadband absorption characteristics [51, 52]; hence, the proposed device can be scaled to operate in the infrared (IR) wavelengths without compromising much of the performance metrics, such as FWHM and ER. So, further, we optimized our designs for the mid-IR spectral range. Since different BOX thicknesses do not significantly affect the simulation results, we kept $t_{BOX} = 3$ μm. Similarly, the same spacer thickness of $t_{HfO_2} = 30$ nm is used to ensure sufficient Fermi-level modulation by the gate voltage. The optical properties of simulated materials were adapted to the mid-IR wavelengths. To shift the operation (resonance) wavelength from ~1580 nm to ~3160 nm we first doubled the thickness of the silicon device layer and the grating layer (namely, for 1D: $t_{Si} = 440$ nm, $h = 180$ nm, for 2D: $t_{Si} = 440$ nm, $h = 380$ nm) and then optimized the grating period $L$ to maximize the ER. Figure 6 shows the simulated ER of mid-IR devices calculated for different $L$. Specifically, we used ER $= 10\log(R_{0.4\,eV}/R_{0.05\,eV})$, where $R_{0.4\,eV}$ and $R_{0.05\,eV}$ are the reflectance amplitudes of resonances obtained for $E_F = 0.4$ eV and $E_F = 0.05$ eV respectively. Using a simple geometrical scaling of doubling $L$ (i.e. $L = 1.6$ μm) we found that the ER is only ~6 dB for both 1D and 2D configurations. Therefore, further optimization is required to maximize ER, which exceeds 20 dB for a slightly modified $L$~1.75 μm and $L$~1.77 μm for 1D and 2D grating, respectively (Fig. 6). By increasing $L$, the resonance wavelength $\lambda_R$ is red-shifted linearly (Fig. 6). Nevertheless, the broad parameter space of $t_{Si}, h$ and $L$ allows for the re-adjusting $\lambda_R$ using the iteration process.

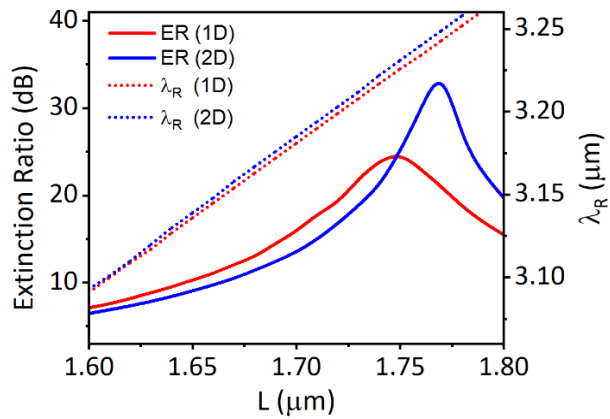

**Fig. 6**. Grating period optimization for operation in mid-IR for 1D and 2D GMR-based active metasurfaces. The left axis shows the extinction ratio, calculated as the ratio between the reflectance amplitudes at resonance $E_F = 0.4$ eV and $E_F = 0.05$ eV. The right axis shows the resonance wavelength obtained for $E_F = 0.4$ eV using different grating periods, which is linearly red-shifted for increased $L$.

Next, using the optimized grating period, we simulated the reflection spectra of polarization-sensitive (TM) and polarization-insensitive (TE/TM) graphene-activated GMR filters operating at mid-IR wavelengths as a function of different Fermi levels (Fig. 7). The Pauli blocking for illumination at ~3.2 μm wavelength is obtained at a lower doping $E_F$~ 0.19 eV compared to the telecom wavelengths, therefore, there is no need to dope graphene to high Fermi levels to bring it to a transparency regime.

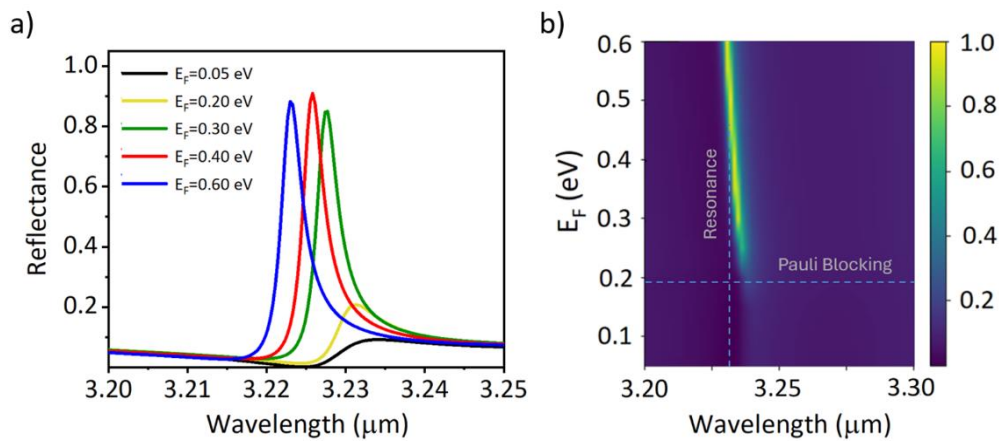

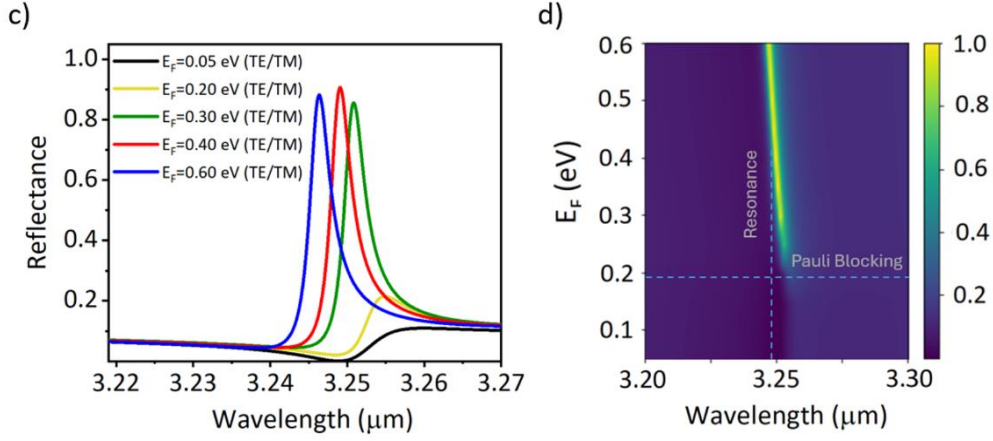

Fig. 7. Reflectance spectra and reflectance map of the device operating at mid-IR wavelengths calculated at different Fermi levels for (a-b) TM polarized 1D grating with $h = 180$ nm, $L = 1.75$ μm, $f = 0.5$ and (c-d) Polarization insensitive 2D grating with $h = 180$ nm, $L = 1.77$ μm, $f = 0.6$. Other parameters for both gratings are set as $t_{HfO_2} = 30$ nm, $t_{Si} = 440$ nm, $t_{BOX} = 3$ μm and $\tau = 0.2$ ps.

After device scaling, the calculated spectral response in the mid-IR is similar to the one presented for telecom wavelengths (Fig. 7 vs. Fig. 2). For the mid-IR operation, we get the $ER > 20$ dB, the resonance linewidth FWHM ~ 3.4 nm scaled with the wavelength and the preserved Q-factor of $Q$~940. The maximum reflectance values and the corresponding resonance wavelength $\lambda_R$ as a function of graphene doping are plotted in Fig. 8a. The peak reflectance reaches above 90% (@ $E_F = 0.4$ eV), and $\lambda_R$ tends to blue-shift at higher doping levels. Figure 8b shows the calculated ER as a function of $E_F$ for the reference resonance at $E_F = 0.4$ eV. The maximum ER ~ 25 dB and ~32 dB were calculated between $E_F = 0.4$ eV and 0.05 eV for 1D and 2D gratings, respectively. Similarly to the telecom wavelengths (1580 nm case, Fig. 5d), the performance parameters FWHM and ER show insignificant variations for the scattering rate in the mid-IR wavelengths, allowing similar performance metrics to be achieved when employing graphene of different quality.

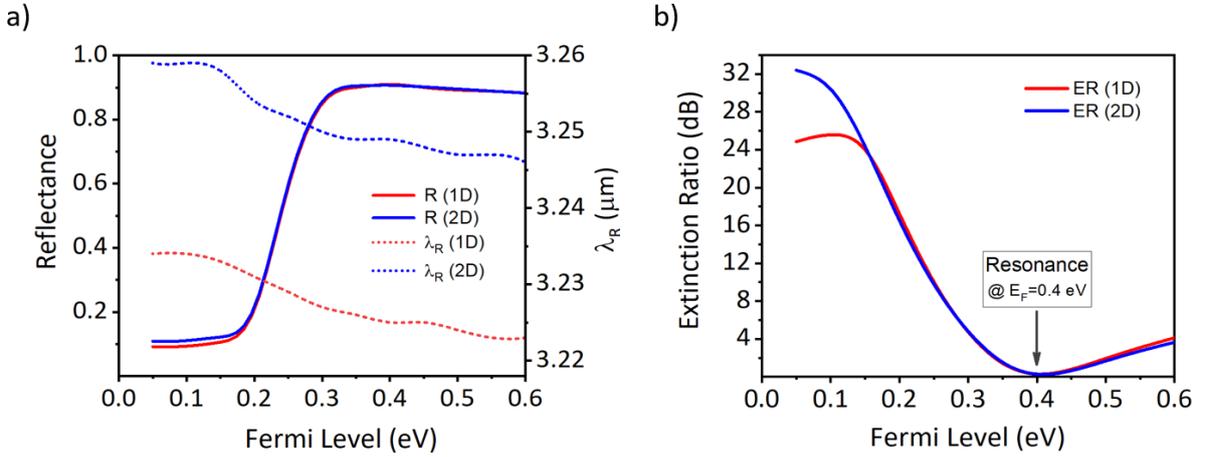

Fig. 8. (a) Maximum reflectance as a function of graphene doping (Fermi level) and corresponding resonance wavelength $\lambda_R$. (b) The extinction ratio ER with respect to the reference resonance at $E_F = 0.4$ eV calculated using different doping levels.

## 4. Tuning by Thermo-Optic Effect

Thermal tuning of the optical refractive index is one of the basic operations in photonics to control phase accumulation of propagating light via the thermo-optic effect, where temperature changes in the materials induce the variation of the effective refractive index of the optical mode (53, 54). Today, microheaters are widely used in photonic integrated circuits (PIC) for thermal stabilization of photonic components, compensation for fabrication tolerances, and spectral tuning and switching of optical resonators (53-55). Recently, atomically thin two-dimensional materials, e.g., semi-metallic graphene (56) and semiconducting transition metal dichalcogenides (TMD), e.g. $MoS_2$ (55), showed highly efficient microheater operation when placed in close proximity (or in direct contact) with integrated photonic devices on-chip (55, 56). In this work, we have discussed so far the electro-optic tuning of the GMR filter by electrostatically modulating the optical properties (loss and refractive index) of SLG that is separated by ~30 nm hafnia gate dielectric from the Si slab. Nonetheless, the same graphene layer, being atomically thin semi-metal, can also conduct an electrical current and generate temperature changes due to the Joule heating, thus modulating the spectral response of the device via the thermo-optic effect. In this scenario, the resistance of the graphene heater can be calculated by:

$$R_G = R_\square \frac{\text{Length}}{\text{Width}} = R_\square \frac{n_x L_x}{n_y L_y} \qquad (4)$$

where $R_\square$ is the SLG sheet resistance, Length $= n_x L_x$ and Width $= n_y L_y$ are the metasurface dimensions, $L_x$ and $L_y$ are the unit cell dimensions in the $x$ and $y$ direction respectively, and $n_x$ and $n_y$ are the integer number of unit cells in $x$ and $y$ direction spanning the total device area. For thermo-optic studies, we considered a rectangular metasurface $n_x = n_y = n$ with a square unit cell $L_x = L_y$ and the total resistance of the graphene heater $R_T = R_\square = 1/(q\mu n_G)$, where $\mu$ and $n_G$ are the charge carriers' mobility and graphene's surface charge density, respectively. For CVD-grown polycrytalline graphene with $\mu \sim 1500$ cm$^2$/Vs at the high doping level $E_F = 0.6$ eV (i.e. transparency, $n_G \sim 2.3 \cdot 10^{13}$ cm$^{-2}$), the calculated $R_T$ is ~160 Ω. Considering the electrical power dissipated per unit cell in the graphene heater is $P_o$, the total power consumption of the full device (for heating) is $P_T = n^2 P_o$ and the corresponding source-drain voltage drop across the graphene layer is $V_{DS} = \sqrt{P_T R_T} = n\sqrt{P_o R_\square}$. For instance, for $P_o = 1$ μW, $L_x = L_y = 1$ μm, $R_T = 160$ Ω and the device area of $A = 10^4$ μm$^2$ (i.e. $n = 100$), we get $V_{DS} \sim 1.25$ V and $P_T \sim 10$ mW.

To study the performance of thermo-optic tuning in our devices, we conducted numerical simulations using finite-element commercial solvers (Lumerical MODE and HEAT) (57). For 2D heat transport simulations, SLG was modelled as a 1 nm thick layer with a thermal conductivity of 600 W/mK [58, 59] and a mesh size of 0.1 nm. Thermal boundary conditions between SLG and $HfO_2$ were modeled using a 1 nm layer of thermal conductivity $0.2 \times 10^{-3}$ W/mK corresponding to the reported thermal boundary conductance of 20 MW m$^{-2}$ K$^{-1}$ for SLG/ $SiO_2$ interfaces [60]. The temperature of the silicon handling wafer was set to 300 K. Figure 9a shows the simulated thermo-optic response of our device as a function of electrical power (per unit length) dissipated in the graphene heater having a unit cell width. We first simulated a steady-state temperature rise $\Delta T$ for different power densities consumed by the heater and calculated the dominant changes of the refractive index in the Si slab based on the thermo-optic effect using $\Delta n_{Si} = (dn/dT_{Si})\Delta T$, where $dn/dT_{Si} = 1.8 \times 10^{-4}$ K$^{-1}$ [61] is the thermo-optic coefficient of silicon at 1550 nm wavelength. Then, the spectral response of the graphene-integrated GMR filter was simulated using finite element optical simulations for SLG doping of $E_F = 0.6$ eV. Due to silicon's positive thermo-optic coefficient, the effective refractive index increases, introducing a phase variation in the filter, resulting in a red shift for increased power densities. The induced spectral shift $\Delta\lambda$ is linear with the increased heating power, as shown in Fig. 9b, following $\Delta\lambda = \frac{\lambda_R}{n_g}\frac{dn}{dT}\Delta T$ [62], where $\frac{dn}{dT}\Delta T$ is the change in the effective refractive index due to the thermo-optic effect, $n_g$ is the group index, and $\lambda_R$ is the resonance wavelength. We noticed a larger spectral shift of the resonance peak due to the thermal effect compared to the electro-optic effect when the graphene doping is modulated to $E_F = 0.5$ eV (Figure 9a). The latter can be understood in terms of a larger overlap of the guided optical mode with the Si slab compared to a nanometer-thick graphene layer.

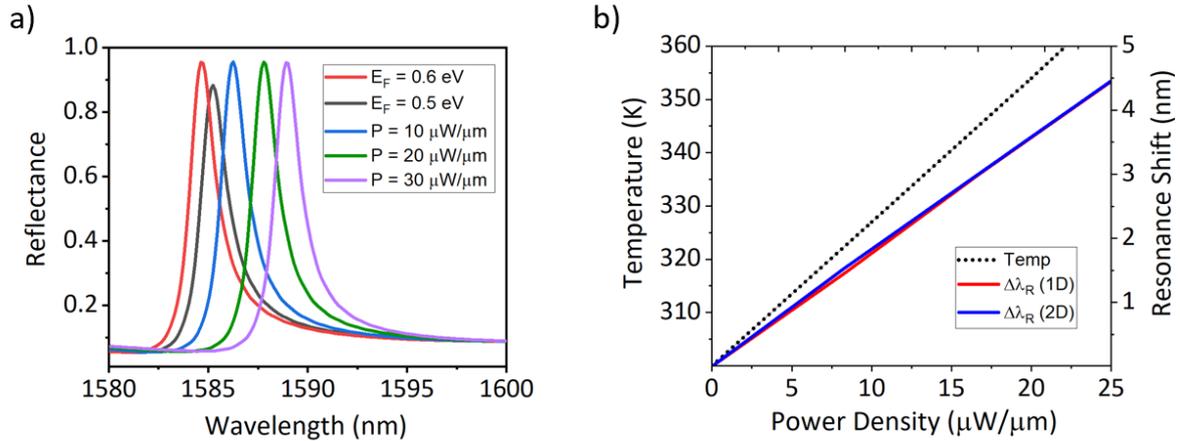

**Fig. 9.** (a) The comparison of resonance tuning using the electro-optic effect of graphene doping modulation and the thermo-optic effect of dissipating the electrical power in graphene for heating. (b) The temperature increase and the resonant wavelength spectral shift as a function of the dissipated heating power.

## 5. Conclusions

In summary, we presented and investigated the advanced CMOS-compatible designs of graphene-integrated silicon GMR-based active metasurfaces. Using the realistic device parameters space, including graphene doping level, gate dielectric breakdown field, and corresponding gate-voltage bias, we demonstrated in simulation that our device can exhibit a high extinction ratio (>25dB) for reflectance modulation, narrow resonance linewidth (~1.5 nm @1550 nm) with a quality factor of Q~1000, and polarization-insensitive operation. Taking advantage of single-layer graphene broadband absorption characteristics, we showed the potential of using the same design considerations for multispectral operation by shifting the device spectral response to the mid-infrared wavelengths using simple geometrical scaling while maintaining performance metrics, such as high extinction ratio, quality factor, and narrow resonance linewidth. Moreover, we showcase that the same device architecture can be employed for thermo-optic tuning using graphene as an integrated microheater, where dual graphene functionality can provide an additional degree of freedom to control the device spectral response and compensate for fabrication tolerances when integrated into more complex photonic systems and functionalities. Our work paves the way for developing advanced broadband silicon-based active metasurfaces for tunable spectral filters and laser mirrors, optical switches, modulators, and sensors.

**Funding.** This work was partially supported by the European Research Council (ERC StG - Multi GRAPH), Israel Science Foundation (Grant No. 1732/18, 2905/20), FLAG-ERA, and Israel Ministry of Science (Grant No. 4490).

**Acknowledgements.** PS greatly acknowledges the Kaufman fellowship.

**Data availability.** Data underlying the results presented in this paper are not publicly available at this time.

**References**
1. Shulin Sun, Qiong He, Jiaming Hao et al., "Electromagnetic metasurfaces: physics and applications," Adv. Opt. Photon. **11**(2), 380-479 (2019).
2. Junghyun Park, Ju-Hyung Kang, Soo Jin Kim, "Dynamic Reflection Phase and Polarization Control in Metasurfaces," *Nano Lett.* **17**(1), 407–413 (2017).
3. Neshev, D., Aharonovich, I., "Optical metasurfaces: new generation building blocks for multi-functional optics," Light Sci Appl **7**, 58 (2018).
4. Li, Aobo, Singh, Shreya and Sievenpiper, Dan. "Metasurfaces and their applications" Nanophotonics **7**(6), 989-1011 (2018).
5. Santonocito A, Patrizi B, Toci G., "Recent Advances in Tunable Metasurfaces and Their Application in Optics," Nanomaterials **13**(10), 1633 (2023).
6. Zehao Wang, Dashuang Liao, Ting Zhang, et al., "Metasurface-based focus-tunable mirror," Opt. Express **27**(21), 30332-30339 (2019).

7. Jingyi Tian, Qiang Li, Jun Lu, and Min Qiu, "Reconfigurable all-dielectric antenna-based metasurface driven by multipolar resonances," Opt. Express **26**(18), 23918-23925 (2018).
8. Tianyou Li, Qunshuo Wei, Bernhard Reineke, *et al.*, "Reconfigurable metasurface hologram by utilizing addressable dynamic pixels," Opt. Express **27**(15), 21153-21162 (2019).
9. P. Berini, "Optical Beam Steering Using Tunable Metasurfaces," ACS Photonics **9**(7), 2204–2218 (2022).
10. Shuyan Zhang, Chi Lok Wong, Shuwen Zeng, "Metasurfaces for biomedical applications: imaging and sensing from a nanophotonics perspective," Nanophotonics, **10**(1), 259-293 (2021).
11. Kazanskiy NL, Khonina SN, Butt MA., "Recent Development in Metasurfaces: A Focus on Sensing Applications," Nanomaterials. **13**(1), 118 (2023).
12. Ping Yu, Jianxiong Li, Na Liu, "Electrically Tunable Optical Metasurfaces for Dynamic Polarization Conversion," Nano Lett. **21**(15), 6690–6695 (2021).
13. Adam C. Overvig, Sander A. Mann, and Andrea Alù," Thermal Metasurfaces: Complete Emission Control by Combining Local and Nonlocal Light-Matter Interactions," Phys. Rev. X **11**, 021050 (2021).
14. Daniil A. Shilkin, Son Tung Ha, Paniagua-Domínguez R, *et al.*, "Ultrafast Modulation of a Nonlocal Semiconductor Metasurface under Spatially Selective Optical Pumping," Nano Lett. **24**(45), 14229–14235 (2024).
15. Abdollahramezani, S., Hemmatyar, O., Taghinejad, M. *et al.,* "Electrically driven reprogrammable phase-change metasurface reaching 80% efficiency," Nat Commun **13**, 1696 (2022).
16. Rosmin Elsa Mohan, Xi Jodi Cheng, Eng Huat Khoo,"Modelling sustainable transparent metasurfaces for tunable near infrared reflectance," Nano-Structures & Nano-Objects **32**, 100924 (2022).
17. Manuel Decker, Christian Kremers, Alexander Minovich, et al., "Electro-optical switching by liquid-crystal controlled metasurfaces," Opt. Express **21**(7), 8879-8885 (2013).
18. P. Vabishchevich *et al.*, "III-V semiconductor metasurface as the optical metamixer," *2018 Conference on Lasers and Electro-Optics (CLEO)*, San Jose, CA, USA, pp. 1-2 (2018).
19. Shuqing Chen, Xinxing Zhou, Yan Luo, *et al*., "Two-Dimensional Material and Metasurface Based Optoelectronics," Adv in Condensed Matter Phys, 8365870 (2019).
20. Osamu Tsutsumi, Tomiki Ikeda, "Photochemical modulation of alignment of liquid crystals and photonic applications," Current Opinion in Solid State and Materials Science **6**(6), 563-568(2002).
21. Patinharekandy Prabhathan, Kandammathe Valiyaveedu Sreekanth, Jinghua Teng *et al.*, "Roadmap for phase change materials in photonics and beyond," iScience **26**(10), 107946 (2023).
22. Wu, P., Pala, R.A., Kafaie Shirmanesh, G. *et al.*, "Dynamic beam steering with all-dielectric electro-optic III–V multiple-quantum-well metasurfaces," *Nat Commun* **10**, 3654 (2019).
23. Vyas, K., Espinosa, D. H. G., Hutama, D., et al., "Group III-V semiconductors as promising nonlinear integrated photonic platforms," Adv. in Phys: X, **7**(1), 2097020 (2022).
24. Zongti Wang, Peng Zhou, Gaige Zheng, "Electrically switchable highly efficient epsilon-near-zero metasurfaces absorber with broadband response," Results in Physics **14**,102376 (2019).
25. Raana Sabri, Ali Forouzmand, and Hossein Mosallaei, "Multi-wavelength voltage-coded metasurface based on indium tin oxide: independently and dynamically controllable near-infrared multi-channels," Opt. Express **28**(3), 3464-3481 (2020).
26. Chang-Hua Liu, You-Chia Chang, Theodore B Norris, *et al.*, "Graphene photodetectors with ultra-broadband and high responsivity at room temperature," Nature Nanotechnology **9**(4), 237-238 (2014).
27. Jin-Ho Lee, Soo-Jeong Park, and Jeong-Woo Choi, "Electrical Property of Graphene and Its Application to Electrochemical Biosensing," Nanomaterials **9**(2), 297 (2019).
28. L A Falkovsky, "Optical properties of graphene," *J. Phys.: Conf. Ser.* **129**, 012004 (2008).
29. Thomas, S., "CMOS-compatible graphene," Nat Electron **1**, 612 (2018).
30. Kai He, Tigang Ning, Jing Li *et al.*, "Wavefront reconfigurable metasurface through graphene micro-ribbons with resonant strategy," Results in Physics **49**, 106484 (2023).
31. Guan Wang, Chen Chen, Ziyang Zhang, "Dynamically tunable infrared grating based on graphene-enabled phase switching of a split ring resonator [Invited]," Optical Materials Express, **9**(1), 56-64 (2019).
32. Jordan A. Goldstein and Dirk R. Englund, "Imaging metasurfaces based on graphene-loaded slot antennas," Opt. Express **29**(2), 1076-1089 (2021).
33. K. Guo, Z. Li and Z. Guo, "Graphene-Integrated Plasmonics Metasurface for Active Controlling Artificial Second Harmonic Generation," in IEEE Access, **8**,159879-159886 (2020).
34. Ferraro, A., Zografopoulos, D.C., Caputo, R. *et al.,* "Guided-mode resonant narrowband terahertz filtering by periodic metallic stripe and patch arrays on cyclo-olefin substrates.," Sci Rep **8**, 17272 (2018)
35. Alok Ghanekar, Rehan Kapadia, Michelle L Povinelli, "Method for tuning absorptivity of a guided-mode resonance grating through period-doubling index perturbation," Journal of Quantitative Spectroscopy and Radiative Transfer **293**, 108367 (2022).
36. Domenico de Ceglia, Marco Gandolfi, Maria Antonietta Vincenti, *et al*., "Transient guided-mode resonance metasurfaces with phase-transition materials," Opt. Lett. **48**(11), 2961-2964 (2023).
37. Lee, K.; LaComb, R.; Britton, B. *et al*., "Silicon-layer guided-mode resonance polarizer with 40-nm bandwidth," IEEE Photonics Technol. Lett., **20**(22), 1857–1859 (2008).
38. Tomohiro Kondo, Shogo Ura, and Robert Magnusson, "Design of guided-mode resonance mirrors for short laser cavities," J. Opt. Soc. Am. A **32**(8), 1454-1458 (2015).
39. S. S. Wang and R. Magnusson, "Theory and applications of guided-mode resonance filters," Appl. Opt. **32**(14), 2606-2613 (1993).
40. Zhaohui Zhang, Yu Zhang, Xinmiao Lu, et al.,"Narrowband and high-transmissivity guided-mode resonance filter in the visible range," Optical Engineering **63**(2), 027102 (2024).
41. Myoung-Seok Kim, Young-Don Ko, Minseong Yun, *et al*., "Characterization and process effects of $HfO_2$ thin films grown by metal-organic molecular beam epitaxy," Materials Science and Engineering: B **123**(1), 20-30 (2005).

42. https://refractiveindex.info/?shelf=main&book=Si&page=Salzberg
43. https://refractiveindex.info/?shelf=main&book=SiO2&page=Malitson
44. A. H. Castro Neto, F. Guinea, N. M. R. Peres, *et al.*, "The electronic properties of graphene," Rev. Mod. Phys. **81(1)**, 109(2009).
45. You-Chia Chang, Chang-Hua Liu, Che-Hung Liu, *et al.*, "Extracting the complex optical conductivity of mono- and bilayer graphene by ellipsometry," Appl. Phys. Lett. **104**(26), 261909 (2014).
46. Domenico de Ceglia, Maria A. Vincenti, Marco Grande, *et al*., "Tuning infrared guided-mode resonances with graphene," J. Opt. Soc. Am. B **33**(3), 426-433 (2016).
47. Geim, A., Novoselov, K.,"The rise of graphene," Nature Mater **6**, 183–191 (2007).
48. Phaedon Avouris, Christos Dimitrakopoulos, "Graphene: synthesis and applications," Materials Today **15**(3), 86-97(2012).
49. Jicheng Wang, Hongyan Shao, Ci Song, *et al*., "Tunable multiple channelled phenomena in graphene-based plasmonic Bragg reflectors," AIP Advances **7**(5), 055204 (2017).
50. CST Studio Suite, "CST Microwave Studio," 2008. http:// www.cst.com
51. Baokun Song, Honggang Gu, Simin Zhu, *et al.* "Broadband optical properties of graphene and HOPG investigated by spectroscopic Mueller matrix ellipsometry," Applied Surface Science **439**, 1079-1087(2018).
52. Wei Li, Guangjun Cheng, Yiran Liang, *et al*., "Broadband optical properties of graphene by spectroscopic ellipsometry," Carbon **99**, 348-353 (2016).
53. Dan-Xia Xu, André Delâge, Pierre Verly, *et al*., "Empirical model for the temperature dependence of silicon refractive index from O to C band based on waveguide measurements," Opt. Express **27**(19), 27229-27241 (2019).
54. Liu, S., Feng, J., Tian, Y. *et al.*, "Thermo-optic phase shifters based on silicon-on-insulator platform: state-of-the-art and a review," Front. Optoelectron. **15**, 9 (2022).
55. Dor Oz, Nathan Suleymanov, Boris Minkovich, *et al.*, "Optically Transparent and Thermally Efficient 2D $MoS_2$ Heaters Integrated with Silicon Microring Resonators," ACS Photonic, **10**(6), 1783–1794 (2023).
56. Junying Li, Yizhong Huang, Yi Song, *et al*., "High-performance graphene-integrated thermo-optic switch: design and experimental validation [Invited]," Opt. Mater. Express **10**(2), 387-396 (2020).
57. Lumerical Inc.
58. Jae Hun Seol et al. "Two-Dimensional Phonon Transport in Supported Graphene", Science 328, 5975 (2010)
59. Zhong Yan, Denis L. Nika, Alexander A. Balandin. "Thermal properties of graphene and few-layer graphene: applications in electronics," IET Circuits, Devices & Systems 9: 4-12 (2015).
60. Cameron J Foss and Zlatan Aksamija. "Quantifying thermal boundary conductance of 2D–3D interfaces, " 2D Materials, Volume 6, Number 2 (2019).
61. J. Komma, C. Schwarz, G. Hofmann, *et al.,* "Thermo-optic coefficient of silicon at 1550 nm and cryogenic temperatures," Appl. Phys. Lett. **101**(4), 041905 (2012).
62. V. Raghunathan, W. N. Ye, J. Hu, *et al*., "Athermal operation of Silicon waveguides: spectral, second order and footprint dependencies," Optics Express **18**(17), 17631-17639 (2010).